\documentclass[11pt]{article}
\usepackage{amssymb}

\usepackage{amsmath}

\newcounter{resultnum}[section]

\newcounter{conclusionnum}[section]

\newcounter{conditionnum}[section]

\newcounter{conjecturenum}[section]

\newcounter{examplenum}[section]

\newcounter{exercisenum}[section]
\newtheorem{lemma}{Lemma}[section]

\newcounter{lemmanum}[section]

\newcounter{notationnum}[section]
\newtheorem{theorem}{Theorem}[section]

\newcounter{theoremnum}[section]
\newtheorem{definition}{Definition}[section]

\newcounter{definitionnum}[section]

\newcounter{corollarynum}[section]
\newtheorem{remark}{Remark}[section]

\newcounter{remarknum}[section]
\newtheorem{proposition}{Proposition}[section]

\newcounter{propositionnum}[section]

\newcounter{acknowledgementnum}[section]

\newcounter{algorithmnum}[section]

\newcounter{axiomnum}[section]

\newcounter{casenum}[section]

\newcounter{claimnum}[section]

\newcounter{summarynum}[section]

\newcounter{problemnum}[section]
\newenvironment{proof}[1][]{\textbf{Proof.} }{}

\begin{document}

\title{Generalized Lagrange Transforms:\\
Finsler Geometry Methods and\\
Deformation Quantization of Gravity}
\date{October 11, 2007}
\author{ Sergiu I. Vacaru\thanks{%
sergiu$_{-}$vacaru@yahoo.com, svacaru@fields.utoronto.ca } \\
{\quad} \\
\textsl{The Fields Institute for Research in Mathematical Science} \\
\textsl{222 College Street, 2d Floor, } \textsl{Toronto \ M5T 3J1, Canada} }
\maketitle

\begin{abstract}
We propose a natural Fedosov type quantization of generalized Lagrange
models and gravity theories with metrics lifted on tangent bundle, or
extended to higher dimension, following some stated geometric/ physical
conditions (for instance, nonholonomic and/or conformal transforms to some
physically important metrics or mapping into a gauge model). Such
generalized Lagrange transforms define canonical nonlinear connection,
metric and linear connection structures and model almost K\"{a}hler
geometries with induced canonical symplectic structure and compatible affine
connection. The constructions are possible due to a synthesis of the
nonlinear connection formalism developed in Finsler and Lagrange geometries
and deformation quantization methods.

\vskip0.3cm \textbf{Keywords:}\ Deformation quantization, quantum gravity,
Finsler and Lagrange geometry, almost K\"{a}hler geometry.

\vskip5pt

MSC:\ 83C45, 81S10, 53D55, 53B40, 53B35, 53D50

PACS:\ 04.60.-m, 02.40.-k, 02.90.+g, 02.40.Yy
\end{abstract}

\newpage

\tableofcontents

\section{ Introduction}

Quantization relates to geometric structures of a classical theory their
algebraic quantum counterparts. This formalism is well elaborated for
geometrized theories in classical mechanics, gauge theory, and various
constructions for almost K\"{a}hler manifolds, symplectic spaces and Poisson
manifolds.

In our recent work \cite{vqgfl}, we elaborated a deformation quantization
formalism for almost K\"{a}hler models of Lagrange and Finsler spaces. For
this article, the review \cite{vrfg} and monograph \cite{vsgg} are the basic
references on nonholonomic manifolds and generalized Lagrange and Finsler
spaces and applications to standard theories of gravity and and
supersymmetric / noncommutative locally anisotropic generalizations. There
is a large class of geometric and physical models working with metrics of
type $g_{ij}(x^{k},y^{a})$ generalizing Lagrange, or Finsler, metrics (when $%
\ ^{L}g_{ij}=(1/2)\partial ^{2}L/\partial y^{i}\partial y^{j},$ induced by a
regular Lagrangian $L(x^{k},y^{a}),$ or $\ ^{F}g_{ij}(x^{k},y^{a}),$ induced
by a fundamental Finsler function $F(x^{k},y^{a}),$ for $L=F^{2}),$ or any
semi--Riemannian metric $g_{ij}(x^{k}),$ for instance, defined as an exact
solution of classical Einstein equations, where $x^{k}$ are local
coordinated on a manifold $M,$ $\dim M=n,$ in the canonical case $n=4,$ and $%
y^{a}$ are fiber/extra dimension coordinates on a corresponding tangent
bundle / higher dimension, denoted $TM/$ $\mathbf{V}^{n+m},\dim TM/\ \mathbf{%
V}^{n+m}=n+n/\ n+m,$ where a generalized model of gravity is constructed.
Such generalized metrics have been constructed as exact solutions in various
models of higher dimension gravity and commutative and noncommutative
versions of the Einstein and gauge gravity, see Refs. \cite{vhep2,vggr,vesnc}%
, see also references in Part II of \cite{vsgg}. They present a substantial
interest in modern physics in order to analyze possible violation of local
Lorentz symmetry by quantum and/or extra dimension effects, see for
instance, Refs. \cite{kost,clmt,brlv} and a recent review in \cite{lehnert}.

One can be defined canonical lifts/ deformations to $TM/$ $\mathbf{V}^{n+m}$
when a metric $g_{ij}(x^{k},y^{a})$ is related to an effective model of
Lagrange, or Finsler, geometry and its almost K\"{a}hler model; the geometry
of generalized Lagrange spaces and their almost Hermitian models are
analyzed in details in Refs. (\cite{ma1987,ma}). In this approach, one
preserves a number of similarities with symplectic models in geometric
mechanics and the almost K\"{a}hler model fundamental geometric objects can
by induced canonically by a generalized Lagrange--Finsler or Einstein metric.

The goal of this paper, is to perform a Fedosov type deformation
quantization of generalized Lagrange metrics induced by nonholonomic
transforms of certain Lagrange, Finsler, or Einstein, metrics to the tangent
bundle or to higher--dimension spacetimes, generalizing the constructions of
our partner work \cite{vqgfl}. We follow the constructions and methods from %
\cite{fedosov1,fedosov2,fedosov,konts1,konts2,karabeg1}. One should be
emphasized that on tangent bundles and higher dimension spacetime such
geometric quantum models are with violation of local Lorentz symmetry by
corresponding quantum/extra dimension effects. Nevertheless, the method of
Fedosov quantization can be re--defined in such a form when the nonholonomic
deformations are performed canonically on semi--Riemannian spaces. This
allows us to elaborate geometric quantum models of the Einstein gravity
without violation of local Lorentz geometry, see considerations in Refs. %
\cite{vqgd} and the first results on Fedosov--Lagrange and Fedosov--Finsler
spaces, and their versions on nonholonomic manifolds in \cite{esv}.

The plan of the paper is as follows:\ In Section 2 we remember some
necessary results on nonlinear connections (N--connection) on tangent
bundles and nonholonomic manifolds provided with N--connection structures.
There are defined effective Lagrange spaces generated by  arbitrary (semi)
Riemannian metrics on base spaces or on tangent  bundles. We prove that such
effective geometric models can be reformulated equivalently as almost K\"{a}%
hler--Lagrange spaces in Section 3. Section 4 is devoted to elaboration of
generalized Fedosov quantization for effective Lagrange spaces. There are
defined the canonical Fedosov operators  and formulated the Fedosov's
theorems for deformation quantization of effective Lagrange spaces (proofs
are similar to those for usual Lagrange and Finsler spaces, see Ref. \cite%
{vqgfl}).

\vskip3pt

\textbf{Conventions:}\ An integral and differential form calculus and a
corresponding tensor analysis in generalized Lagrange--Finsler spaces
require a more sophisticate system of denotations, see details in \cite%
{vrfg,vqgfl}. We shall use Einstein's summation convention in local
expressions. \vskip3pt

\textbf{Acknowledgments:}\ This work is devoted to Academician Prof. Dr.
Radu Miron at his 80th anniversary.

\section{Effective Lagrange Spaces}

Let us denote by $V^{2n},$ $\dim V^{2n}=2n,$ where $n\geq 2,$ a manifold
(space) of necessary smooth class (for instance, $V^{2n}=TM$ is a tangent
bundle of a manifold $V^{n},$ or $V^{2n}$ is Riemann--Cartan manifold, in
general, enabled with nonlinear connection structure \cite{vrfg,vsgg}).%
\footnote{%
for constructions related to the Einstein gravity, in this work, $n=4$} We
label the local coordinates on $V^{2n}$ in the form $u^{\alpha
}=(x^{i},y^{a}),$ or $u=(x,y),$ where indices run values $i,j,...=1,2,...n$
and $a,b,...=n+1,n+2,...,n+2;$ for $TM,$ we can use the same values for
indices of base coordinates $x^{i}$ and fiber coordinates $y^{i}$ and for
certain geometric constructions such conventional horizontal (h) and
vertical (v) can be contracted.

\subsection{N--anholonomic manifolds and tangent bundles}

A nonlinear connection (N--connection) $\mathbf{N}$ on a space $T(V^{2n})$
can be defined by a Whitney sum (nonholonomic distribution)%
\begin{equation}
T(V^{2n})=h(V^{2n})\oplus v(V^{2n}),  \label{whitney}
\end{equation}%
into conventional h-- and v--subspaces, given locally to by a set of
coefficients $N_{i}^{a}(x,y)$ defined with respect to a coordinate basis $%
\partial _{\alpha }=\partial /\partial u^{\alpha }=(\partial _{i}=\partial
/\partial x^{i},\partial _{a}=\partial /\partial y^{a})$ and its dual $%
du^{\beta }=(dx^{j},dy^{b}).$ The subclass of linear connections consists
from a particular case when $N_{i}^{a}=\Gamma _{b}^{a}(x)y^{b}.$ The
curvature of a N--connection is defined by the corresponding Neijenhuis
tensor
\begin{equation*}
\Omega _{ij}^{a}=\frac{\partial N_{i}^{a}}{\partial x^{j}}-\frac{\partial
N_{j}^{a}}{\partial x^{i}}+N_{i}^{b}\frac{\partial N_{j}^{a}}{\partial y^{b}}%
-N_{j}^{b}\frac{\partial N_{i}^{a}}{\partial y^{b}}.
\end{equation*}%
In general, in this work, the spacetimes will be modelled as N--anholonomic
manifolds $\mathbf{V}^{2n},$ i.e. as manifolds, or tangent/vector bundles,
provided nonholonomic distributions defining N--connection structures.

On $\mathbf{V}^{2n},$ there are defined certain preferred frame structures
(depending linearly on $N_{j}^{a}$) denoted $\mathbf{e}_{\nu }=(\mathbf{e}%
_{i},e_{a}),$ where
\begin{equation}
\mathbf{e}_{i}=\frac{\partial }{\partial x^{i}}-N_{i}^{a}(u)\frac{\partial }{%
\partial y^{a}}\mbox{ and
}e_{a}=\frac{\partial }{\partial y^{a}},  \label{dder}
\end{equation}%
and the dual frame (coframe) structure is $\mathbf{e}^{\mu }=(e^{i},\mathbf{e%
}^{a}),$ where
\begin{equation}
e^{i}=dx^{i}\mbox{ and }\mathbf{e}^{a}=dy^{a}+N_{i}^{a}(u)dx^{i},
\label{ddif}
\end{equation}%
satisfying nontrivial nonholonomy relations
\begin{equation}
\lbrack \mathbf{e}_{\alpha },\mathbf{e}_{\beta }]=\mathbf{e}_{\alpha }%
\mathbf{e}_{\beta }-\mathbf{e}_{\beta }\mathbf{e}_{\alpha }=W_{\alpha \beta
}^{\gamma }\mathbf{e}_{\gamma }  \label{anhrel}
\end{equation}%
with (antisymmetric) nontrivial anholonomy coefficients $W_{ia}^{b}=\partial
_{a}N_{i}^{b}$ and $W_{ji}^{a}=\Omega _{ij}^{a}.$\footnote{%
We use boldface symbols for spaces (geometric objects) provided with
N--connection structure (adapted to the h-- and v--splitting defined by (\ref%
{whitney}), we say that they are N--adapted). The N--adapted tensors,
vectors, forms... are called respectively distinguished tensors,..., in
brief, d--tensors, d--vectors, d--forms.}

\begin{definition}
A space, i.e. a tangent bundle, manifold,..., $\mathbf{V}^{2n}$ is called
nonholonomic (N--anholonomic) if it is enabled with a nonholonomic
distribution (N--connection).
\end{definition}

In this work, the geometric constructions will be adapted to the
N--connec\-ti\-on structure (\ref{whitney}).

\begin{definition}
A (h,v)--metric on $\mathbf{V}^{2n}$ is a symmetric d--tensor field
\begin{equation}
\mathbf{g}=g_{ij}(x,y)\ e^{i}\otimes e^{j}+h_{ab}(x,y)\ \mathbf{e}%
^{a}\otimes \ \mathbf{e}^{b},  \label{hvmetr}
\end{equation}%
where the first / second term defines the h--metric / v--metric.
\end{definition}

From the class of affine connections on $\mathbf{V}^{2n},$ one prefers to
work with N--adapted linear connections:

\begin{definition}
A d--connection $\mathbf{D}=(hD;vD),$ with local coefficients computed with
respect to (\ref{dder}) and (\ref{ddif}), $\{\mathbf{\Gamma }_{\beta \gamma
}^{\alpha }=(L_{jk}^{i},L_{bk}^{a};C_{jc}^{i},C_{bc}^{a})\},$  preserves
under parallel transports the distribution (\ref{whitney}).
\end{definition}

One defines respectively the nonmetricity field, torsion and curvature of a
d--connection $\mathbf{D,}$
\begin{eqnarray}
\mathbf{Q}_{\mathbf{X}} &\doteqdot &\mathbf{D}_{\mathbf{X}}\mathbf{g},
\label{nm} \\
\mathbf{T}(\mathbf{X},\mathbf{Y}) &\doteqdot &\mathbf{D}_{\mathbf{X}}\mathbf{%
Y}-\mathbf{D}_{\mathbf{Y}}\mathbf{X}-[\mathbf{X},\mathbf{Y}],  \label{ators}
\\
\mathbf{R}(\mathbf{X},\mathbf{Y})\mathbf{Z} &\doteqdot &\mathbf{D}_{\mathbf{X%
}}\mathbf{D}_{\mathbf{Y}}\mathbf{Z}-\mathbf{D}_{\mathbf{Y}}\mathbf{D}_{%
\mathbf{X}}\mathbf{Z}-\mathbf{D}_{[\mathbf{X},\mathbf{Y}]}\mathbf{Z},
\label{acurv}
\end{eqnarray}%
where the symbol ''$\doteqdot $'' states ''by definition'' and $[\mathbf{X},%
\mathbf{Y}]\doteqdot \mathbf{XY}-\mathbf{YX,}$ for any d--vectors $\mathbf{X}
$ and $\mathbf{Y}.$ With respect to fixed local bases $\mathbf{e}_{\alpha }$
and $\mathbf{e}^{\beta },$ the coefficients $\mathbf{Q}=\{\mathbf{Q}_{\alpha
\beta \gamma }=\mathbf{D}_{\alpha }\mathbf{g}_{\beta \gamma }\},\mathbf{T}=\{%
\mathbf{T}_{\ \beta \gamma }^{\alpha }\}$ and $\mathbf{R}=\{\mathbf{R}_{\
\beta \gamma \tau }^{\alpha }\}$ can be computed by introducing $\mathbf{X}%
\rightarrow \mathbf{e}_{\alpha },\mathbf{Y}\rightarrow \mathbf{e}_{\beta }%
\mathbf{,Z}\rightarrow \mathbf{e}_{\gamma }$ into respective formulas (\ref%
{nm}), (\ref{ators}) and (\ref{acurv}).

In gravity theories, one uses three others important geometric objects: the
Ricci tensor, $Ric(\mathbf{D})=\{\mathbf{R}_{\ \beta \gamma }\doteqdot
\mathbf{R}_{\ \beta \gamma \alpha }^{\alpha }\},$ the scalar curvature, $\
^{s}R\doteqdot \mathbf{g}^{\alpha \beta }\mathbf{R}_{\alpha \beta }$ ($%
\mathbf{g}^{\alpha \beta }$ being the inverse matrix to $\mathbf{g}_{\alpha
\beta }),$ and the Einstein tensor, $\mathbf{E}=\{\mathbf{E}_{\alpha \beta
}\doteqdot \mathbf{R}_{\alpha \beta }-\frac{1}{2}\mathbf{g}_{\alpha \beta }\
^{s}R\}.$

In the canonical geometric models on $\mathbf{V}^{2n},$ one works with
metrical d--connections for which $\mathbf{D}_{\mathbf{X}}\mathbf{g}=0.$
Following the Kawaguchi--Miron methods (\cite{ma1987,ma,vsgg}), any
d--connection (in general, with nonmetricity) can be transformed into
metrical ones, which, in their turns, can be reduced to the canonical
d--connection.

For bundle spaces and manifolds of dimension $2n,$ enabled with
N--connec\-ti\-on structure $N_{j}^{a},$ it is preferred to work with the
so--called canonical d--connection $\widehat{\mathbf{D}}=(h\widehat{D};v%
\widehat{D})=\{\widehat{\mathbf{\Gamma }}_{\beta \gamma }^{\alpha }\}$ which
is metric compatible, $\widehat{\mathbf{D}}_{\mathbf{X}}\mathbf{g}=0$ and
completely defined by $h\widehat{D}=\{\widehat{L}_{jk}^{i}\}$ and $v\widehat{%
D}=\{\widehat{C}_{jc}^{i}\},$\footnote{%
It should be noted that on spaces of arbitrary dimension $n+m,m\neq n,$ we
have $\mathbf{D}=(hD;vD)=\{\mathbf{\Gamma }_{\beta \gamma }^{\alpha
}=(L_{jk}^{i},L_{bk}^{a};C_{jc}^{i},C_{bc}^{a})\};$ in such cases we can not
identify $L_{jk}^{i}$ with $L_{bk}^{a}$ and $C_{jc}^{i}$ with $C_{bc}^{a}.$
On $TM,$ the h- and v--indices \ can be identified \ in the form ''$%
i\rightarrow a",$ where $i=1=a=n+1,$ $i=2=a=n+2,$ ... $i=n=a=n+n.$} where
with respect to the N--adapted bases (\ref{dder}) and (\ref{ddif}),
\begin{equation}
\widehat{L}_{\ jk}^{i}=\frac{1}{2}g^{ih}(\mathbf{e}_{k}g_{jh}+\mathbf{e}%
_{j}g_{kh}-\mathbf{e}_{h}g_{jk}),\ \widehat{C}_{\ bc}^{a}=\frac{1}{2}%
h^{ae}(e_{b}h_{ec}+e_{c}h_{eb}-e_{e}h_{bc}).  \label{cdcc}
\end{equation}%
This connection is uniquely defined by the metric structure\ (\ref{hvmetr})
to satisfy the conditions $\widehat{T}_{jk}^{i}=0$ and $\widehat{T}%
_{bc}^{a}=0$ (when torsion vanishes in the h-- and v--subspaces) but
nevertheless the torsion $\widehat{\mathbf{T}}_{\ \beta \gamma }^{\alpha }$
has nontrivial torsion components
\begin{equation}
\widehat{T}_{jc}^{i}=\widehat{C}_{\ jc}^{i},\widehat{T}_{ij}^{a}=\Omega
_{ij}^{a},\widehat{T}_{ib}^{a}=e_{b}N_{i}^{a}-\widehat{L}_{\ bi}^{a},
\label{cdtors}
\end{equation}%
which are induced by respective nonholonomic deformations and also
completely defined by the N--connection and d--metric coefficients. The
N--adapted coefficients of curvature $\widehat{\mathbf{R}}_{\ \beta \gamma
\tau }^{\alpha }$ of $\widehat{\mathbf{D}}$ are computed
\begin{eqnarray}
\widehat{R}_{\ hjk}^{i} &=&\mathbf{e}_{k}\widehat{L}_{\ hj}^{i}-\mathbf{e}%
_{j}\widehat{L}_{\ hk}^{i}+\widehat{L}_{\ hj}^{m}\widehat{L}_{\ mk}^{i}-%
\widehat{L}_{\ hk}^{m}\widehat{L}_{\ mj}^{i}-\widehat{C}_{\ ha}^{i}\Omega
_{\ kj}^{a},  \label{cdcurv} \\
\widehat{P}_{\ jka}^{i} &=&e_{a}\widehat{L}_{\ jk}^{i}-\widehat{\mathbf{D}}%
_{k}\widehat{C}_{\ ja}^{i},\ \widehat{S}_{\ bcd}^{a}=e_{d}\widehat{C}_{\
bc}^{a}-e_{c}\widehat{C}_{\ bd}^{a}+\widehat{C}_{\ bc}^{e}\widehat{C}_{\
ed}^{a}-\widehat{C}_{\ bd}^{e}\widehat{C}_{\ ec}^{a}.  \notag
\end{eqnarray}

It should be noted that geometrically the d--connection $\widehat{\mathbf{D}}
$ is different from the well known Levi Civita connection $\nabla =\{\
^{\shortmid }\Gamma _{\beta \gamma }^{\alpha }\},$ uniquely defined by the
same metric (\ref{hvmetr}) in order to satisfy the conditions $\nabla
_{\alpha }\mathbf{g}_{\beta \gamma }=0$ and $\ ^{\shortmid }T_{\ \beta
\gamma }^{\alpha }=0.$ This linear connection is not a d--connection because
it does not preserve under parallelism the conditions (\ref{whitney}).
Nevertheless, we can work equivalently with both connections $\nabla $ and $%
\widehat{\mathbf{D}},$ because they are defined in unique forms for the same
metric (even these connection have different geometric and physical
meaning).  All data computed for one connection can be recomputed for
another one by using the distorsion tensor, $\mathbf{Z}_{\beta \gamma
}^{\alpha }$ when $\widehat{\mathbf{\Gamma }}_{\beta \gamma }^{\alpha }=\
^{\shortmid }\Gamma _{\beta \gamma }^{\alpha }+\mathbf{\ Z}_{\beta \gamma
}^{\alpha }.$ We conclude that a (semi) Riemannian N--anholonomic manifold
provided with metric $\mathbf{g}$ and N--connection $\mathbf{N}$ structures
can be equivalently described as a usual Riemannian manifold enabled with
the connection $\nabla (\mathbf{g})$ (in a form not adapted to
N--connection) or as a Riemann--Cartan manifold with the torsion $\widehat{%
\mathbf{T}}(\mathbf{g})$ induced canonically by $\mathbf{g}$ and $\mathbf{N}$
(adapted to the N--connection structure).

\subsection{Effective Lagrange spaces and (semi) Riemannian metrics}

There is a class of N--anholonomic Riemann manifolds $\mathbf{V}^{2n}$ which
can be equivalently modelled as effective Lagrange and, as a consequence, as
effective almost K\"{a}hler manifolds, see explicit examples in Refs. \cite%
{ma1987,ma,vrfg}. Let us introduce an effective Lagrange function $\mathcal{L%
}$ as a functional of an arbitrary d--metric $g_{ij}=h_{ij}(x,y),$ arbitrary
functions on $(x,y),$ frame transforms in v--space and variables $y^{a},$%
\begin{equation}
\mathcal{L=L}(h_{ij}(x,y),y^{a},e_{\ a^{\prime }}^{a}(x,y),f(x,y),...).
\label{funct}
\end{equation}%
In Refs. \cite{ma1987,ma}, $\mathcal{L}=h_{ab}(x,y)y^{a}y^{b}$ is called the
absolute energy. There were performed constructions with nonholonomically
transformed $h_{a^{\prime }b^{\prime }}(x,y)=e_{\ a^{\prime }}^{a}(x,y)e_{\
b^{\prime }}^{b}(x,y)h_{ab}(x,y),$ in order to generate constant coefficient
Hessians $\frac{1}{2}\frac{\partial ^{2}\mathcal{L}}{\partial y^{a^{\prime
}}\partial y^{b^{\prime }}}=const$ which allowed to define bi--Hamilton and
solitonic hierarchies for (semi) Riemannian and Lagrange--Finsler metrics %
\cite{vrf03}, or with conformal lifts of a metric $g_{ab}(x)$ from $M$ to $%
TM,$ when $\mathcal{L}=f(x,y)g_{ab}(x)y^{a}y^{b},$ see \cite{vrfg}. We note
that in certain particular cases, the nontrivial coefficients $h_{ij}(x)$
(or $h_{ij}(x,y))$ in a functional $\mathcal{L}$, can define an exact
solution of the Einstein equations on a manifold $M$ (or $\mathbf{V}^{2n}$).
For our purposes, it is important to use any ''simple'' geometric or
physical principle to construct a functional $\mathcal{L}$, for instance,
following some classes of conformal transforms, quadratic dependence on
variable $y^{a},$ frame transforms, ... \footnote{%
The explicit form of $\mathcal{L}$ depends on the type of geometric/
physical model one wonts to construct on $\mathcal{L}.$ For simplicity, in
this work, we shall work with a a general class of so--called regular
functionals not fixing an explicit theory on tangent bundle or extra
dimension spacetime.}

\begin{definition}
An effective Lagrange v--metric
\begin{equation}
^{\mathcal{L}}g_{ij}=(1/2)\partial ^{2}\mathcal{L}/\partial y^{i}\partial
y^{j}  \label{hessel}
\end{equation}%
is regular if $\det |^{\mathcal{L}}g_{ij}|\neq 0,$ i.e. this metric is
nondegenerate. We call an effective Lagrange function $\mathcal{L}$ to be
regular if its v--metric $^{\mathcal{L}}g_{ij}$ is regular.
\end{definition}

In general, a metric $h_{ij}$ in (\ref{funct}) is not a Lagrange, or
Finsler, type, i.e. $h_{ij}$ can not be represented in the form $%
(1/2)\partial ^{2}L/\partial y^{i}\partial y^{j},$ or $(1/2)\partial
^{2}F^{2}/\partial y^{i}\partial y^{j},$ where $L(x,y)$ is a regular
Lagrangian, or $L=F^{2}(x,y)$ is defined by a fundamental Finsler function $%
F(x,\lambda y)=|\lambda |F(x,\lambda y).$ For instance, a metric of type $%
h_{ij}=e^{2\gamma (x,y)}\ ^{E}g_{ij}(x),$ where $\ ^{E}g_{ij}(x)$ is a
solution of the Einstein equations, conformally deformed from $M$ to $TM,$
defines a generalized Lagrange space, but not a Lagrange space.
Nevertheless, for a regular $\mathcal{L}$ chosen for a theory on $\mathbf{V}%
^{2n},$ we can model an effective Lagrange space and perform a deformation
quantization as we considered in Ref. \cite{vqgfl}.

One hold the results:

\begin{theorem}
For a regular effective Lagrangian $\mathcal{L},$ any metric $h_{ij}$
induces a geometrical model with fundamental geometric structures and
objects:

\begin{enumerate}
\item a canonical N--connection $\ $%
\begin{eqnarray}
\ ^{\mathcal{L}}N_{j}^{a} &=&\frac{\partial \ ^{\mathcal{L}}G^{a}(x,y)}{%
\partial y^{j}},  \label{cncl} \\
\ ^{\mathcal{L}}G^{k} &=&\frac{1}{4}\ ^{\mathcal{L}}g^{kj}\left( y^{i}\frac{%
\partial ^{2}\mathcal{L}}{\partial y^{j}\partial x^{i}}-\frac{\partial
\mathcal{L}}{\partial x^{j}}\right) ,  \notag
\end{eqnarray}%
where $\ ^{\mathcal{L}}g^{kj}$ is inverse to $^{\mathcal{L}}g_{ij}$ (\ref%
{hessel}), inducing preferred frame structures $\ ^{\mathcal{L}}\mathbf{e}%
_{\nu }=(\ ^{\mathcal{L}}\mathbf{e}_{i},e_{a})$ (\ref{dder}) and $\ ^{%
\mathcal{L}}\mathbf{e}^{\mu }=(e^{i},\ ^{\mathcal{L}}\mathbf{e}^{a})$ (\ref%
{ddif}), with $\ N_{j}^{a}=\ ^{\mathcal{L}}N_{j}^{a};$

\item a canonical (h,v)--metric with $^{\mathcal{L}}g_{ij}=\ ^{\mathcal{L}%
}h_{ij},$
\begin{equation}
\ ^{\mathcal{L}}\mathbf{g}=\ ^{\mathcal{L}}g_{ij}(x,y)\ e^{i}\otimes e^{j}+\
^{\mathcal{L}}g_{ab}(x,y)\ \ ^{\mathcal{L}}\mathbf{e}^{a}\otimes \ ^{%
\mathcal{L}}\mathbf{e}^{b};  \label{hvmgl}
\end{equation}

\item a canonical d--connection $\ ^{\mathcal{L}}\widehat{\mathbf{D}}=\{\ ^{%
\mathcal{L}}\widehat{\mathbf{\Gamma }}_{\beta \gamma }^{\alpha }\}=(h\ ^{%
\mathcal{L}}\widehat{D}=\{\ ^{\mathcal{L}}\widehat{L}_{jk}^{i}\};v\ ^{%
\mathcal{L}}\widehat{D}=\{\ ^{\mathcal{L}}\widehat{C}_{jc}^{i}\})$ with the
coefficients computed respectively by formulas (\ref{cdcc}) with $g_{ij}$
and $h_{ij}$ replaced by$\ ^{\mathcal{L}}g_{ij}.$
\end{enumerate}
\end{theorem}

\begin{proof}
The most important motivation for such constructions is the fact that the
Euler--Lagrange equations, $\frac{d}{d\tau }\frac{\partial \mathcal{L}}{%
\partial y^{i}}-\frac{\partial \mathcal{L}}{\partial x^{i}}=0, $ where $\tau
$ is the variation parameter, are equivalent to the ''nonlinear geodes\-ic''
(equivalently, semi--spray) equations
\begin{equation*}
\frac{d^{2}x^{k}}{d\tau ^{2}}+2\ ^{\mathcal{L}}G^{k}(x,y)=0.
\end{equation*}%
In result, we can define the canonical N--connection $\ ^{\mathcal{L}%
}N_{j}^{a}$ (\ref{cncl}), which for a chosen functional $\mathcal{L}$ is
defined by a d--tensor field $h_{ij}$ and state a preferred frame and
co--frame structure $^{\mathcal{L}}\mathbf{e}_{\nu }$ and $^{\mathcal{L}}%
\mathbf{e}^{\mu }.$ The d--metric (\ref{hvmgl}) is an example of Sasaki lift
when the coefficients are $\ ^{\mathcal{L}}g_{ij},$ which in its turn
defines the canonical d--connection $^{\mathcal{L}}\widehat{\mathbf{D}}.$ $%
\Box $
\end{proof}

One follows:

\begin{remark}
For $^{\mathcal{L}}\widehat{\mathbf{D}},$ the nontrivial coefficients of
torsion $\ ^{\mathcal{L}}\widehat{\mathbf{T}}_{\ \beta \gamma }^{\alpha }$
and curvature $\ ^{\mathcal{L}}\widehat{\mathbf{R}}_{\ \beta \gamma \tau
}^{\alpha }$ are computed by introducing the nontrivial coefficients of $^{%
\mathcal{L}}\widehat{\mathbf{D}}$ and $\ ^{\mathcal{L}}N_{j}^{a},$
respectively, into formulas (\ref{cdtors}) and (\ref{cdcurv}).
\end{remark}

We conclude that for any chosen regular functional $\mathcal{L}$ any
d--tensor field $h_{ij}$ induces an effective Lagrange space enabled with
canonical $\ ^{\mathcal{L}}\mathbf{N,}\ ^{\mathcal{L}}\mathbf{g} $ and $^{%
\mathcal{L}}\widehat{\mathbf{D}}$ structures.

\section{Almost K\"{a}hler -- Generalized Lagrange Spac\-es}

The aim of this section is to prove that any d--tensor $h_{ij}$ for a fixed
regular effective Lagrange structure $\mathcal{L}$ also defines canonically
a nonholonomic almost K\"{a}hler structure on $\mathbf{V}^{2n}.$

We define an almost K\"{a}hler structure using the preferred N--adapted
frames, i.e. the canonical N--connection $\ ^{\mathcal{L}}N_{j}^{a}$ (\ref%
{cncl}) (in this paper, we use the constructions from \cite{ma1987,ma}
generalized for arbitrary functionals $\mathcal{L}$ \cite{vrfg}). The almost
complex structure is stated by a linear operator $\mathbf{J}$ acting on the
vectors following formulas $\mathbf{J}(\ ^{\mathcal{L}}\mathbf{e}_{i})=-e_{i}
$ and $\mathbf{J}(e_{i})=\ ^{\mathcal{L}}\mathbf{e}_{i},$ where the
superposition $\mathbf{J\circ J=-I,}$ for $\mathbf{I}$ being the unity
matrix. By straightforward computations one prove:

\begin{proposition}
\label{pr0}A regular effective Lagrangian $\mathcal{L}(x,y)$ induces a
canonical 1--form $ \ ^{\mathcal{L}}\omega =\frac{1}{2}\frac{\partial
\mathcal{L}}{\partial y^{i}}e^{i}$ and the metric $\ ^{\mathcal{L}}\mathbf{g}
$ (\ref{hvmgl}) induces a canonical 2--form
\begin{equation}
\ ^{\mathcal{L}}\mathbf{\theta }=\ ^{\mathcal{L}}g_{ij}(x,y)\ ^{\mathcal{L}}%
\mathbf{e}^{i}\wedge e^{j}.  \label{asstr}
\end{equation}%
associated to $\mathbf{J}$ following formulas $\ ^{\mathcal{L}}\mathbf{%
\theta (X,Y)}\doteqdot \ ^{\mathcal{L}}\mathbf{g}\left( \mathbf{JX,Y}\right)
$ for any vectors $\mathbf{X}$ and $\mathbf{Y}$ on $TM$ decomposed with
respect to the adapted to $^{\mathcal{L}}N_{j}^{a}$ basis (\ref{dder}).
\end{proposition}

One holds:

\begin{theorem}
\label{th0}Any d--tensor field $h_{ij}(x,y)$ for a regular effective
Lagrangian $\mathcal{L}$ induces an almost K\"{a}hler structure for which $%
d\ ^{\mathcal{L}}\mathbf{\theta =}\ ^{\mathcal{L}}\mathbf{\omega }=0.$
\end{theorem}

\begin{proof}
It follows from Proposition \ref{pr0}; computations are similar to those for
usual Lagrange spaces provided in \cite{ma1987,ma}. $\Box $
\end{proof}

\begin{definition}
An almost K\"{a}hler d--connection $\ ^{\theta }\mathbf{D}$ is compatible
both with the almost K\"{a}hler $\left( \ ^{\mathcal{L}}\mathbf{\theta ,J}%
\right) $ and canonical N--connection structures $\ ^{\mathcal{L}}\mathbf{N}$
and satisfies the conditions $\ ^{\theta }\mathbf{D}_{\mathbf{X}}\ ^{%
\mathcal{L}}\mathbf{g}=0$  and $\ ^{\theta }\mathbf{D}_{\mathbf{X}}\mathbf{J}%
=0, $  for any vector $\mathbf{X}=X^{i}\mathbf{e}_{i}+X^{a}e_{a}.$
\end{definition}

By a straightforward computation one proves:

\begin{theorem}
\label{th1}The canonical d--connection $^{\mathcal{L}}\widehat{\mathbf{D}}$
defines also a unique canonical almost K\"{a}hler d--connection $^{\theta }%
\widehat{\mathbf{D}}=\ ^{\mathcal{L}}\widehat{\mathbf{D}}$ for which with
respect to N--adapted frames the coefficients $\widehat{T}_{jk}^{i}=0$ and $%
\widehat{T}_{bc}^{a}=0.$
\end{theorem}

We get a K\"{a}hlerian model if the respective almost complex structure $%
\mathbf{J}$ is integrable.

As a matter of principle, for any d--tensor field $h_{ij}(x,y),$ we can
consider a metric (\ref{hvmetr}) with $h_{ij}(x,y)=g_{ij}(x,y)$ for any
prescribed N--connection structure $N_{i}^{a}(x,y).$ Following formulas (\ref%
{cdcc}), for $g_{ij}=h_{ij},$ we compute the canonical d--connection $%
\widehat{\mathbf{D}}=\{\widehat{L}_{jk}^{i},\widehat{C}_{jc}^{i}\}.$ In this
case, we also can define an almost complex and symplectic structure as in
Proposition \ref{pr0} stating that $\ \mathbf{\theta (X,Y)}\doteqdot \
\mathbf{g}\left( \mathbf{JX,Y}\right) $ and correspondingly construct the
canonical almost K\"{a}hler d--connection $^{\theta }\widehat{\mathbf{D}}=\
\widehat{\mathbf{D}}$ as in Theorem \ref{th1}. This allows us to model a
generalized Lagrange spaces as an almost Hermitian geometry, because in this
case $d\ \mathbf{\theta \neq 0,}$ i.e. this form is not closed. A model of
geometric quantization can be elaborated in the "simplest" way if we work
with closed symplectic forms, which is possible if we use $h_{ij}(x,y)$ in
order to define certain canonical constructions resulting in an almost K\"{a}%
hler structure and not in an almost Hermitian one. That why we considered
effective Lagrange spaces derived for a given $h_{ij}(x,y)$ and regular
effective Lagrange function $\mathcal{L}:$ this way, one obtains an
effective but closed $\ ^{\mathcal{L}}\mathbf{\theta .}$

\section{Generalized Fedosov--Lagrange Quanti\-za\-ti\-on}

We formulate a Fedosov type deformation quantization \cite%
{fedosov1,fedosov2,fedosov,karabeg1} for any d--tensor field $h_{ij}(x,y)$
on $\mathbf{V}^{2n}$ which for a regular functional $\mathcal{L}(x,y)$
defines a canonical effective K\"{a}hler--Lagrange geometry. Proofs of
results will be omitted because they are completely similar to those for
Lagrange (Finsler) spaces presents in Ref. \cite{vqgfl} (the corresponding
geometric objects labelled, for instance, as $^{L}\mathbf{\theta ,}^{L}%
\widehat{\mathbf{D}},$ ... should be changed into $^{\mathcal{L}}\mathbf{%
\theta ,}$ $^{\mathcal{L}}\widehat{\mathbf{D}},...$ which allows to perform
similar constructions for effective Lagrange spaces; this emphasizes the
crucial importance of the methods of Finsler and Lagrange geometry in order
to perform deformation quantization of arbitrary d--tensor structures, see
discussions in the mentioned work).

\subsection{Effective Fedosov operators}

We denote by $C^{\infty }(V)[[v]]$ the spaces of formal series in variable $%
v $ with coefficients from $C^{\infty }(V)$ on a Poisson manifold $%
(V,\{\cdot ,\cdot \}).$ An associative algebra structure on $C^{\infty
}(V)[[v]]$ with a $v$--linear and $v$--adically continuous star product
\begin{equation}
\ ^{1}f\ast \ ^{2}f=\sum\limits_{r=0}^{\infty }\ _{r}C(\ ^{1}f,\ ^{2}f)\
v^{r},  \label{starp}
\end{equation}%
where $\ _{r}C,r\geq 0,$ are bilinear operators on $C^{\infty }(V)$ with $\
_{0}C(\ ^{1}f,\ ^{2}f)=\ ^{1}f\ ^{2}f$ and $\ _{1}C(\ ^{1}f,\ ^{2}f)-\
_{1}C(\ ^{2}f,\ ^{1}f)=i\{\ ^{1}f,\ ^{2}f\},$ with $i$ being the complex
unity.

On $\mathbf{V}^{2n}$ enabled with canonical effective Lagrange structures,
we introduce the tensor $\ ^{\mathcal{L}}\mathbf{\Lambda }^{\alpha \beta
}\doteqdot \ ^{\mathcal{L}}\theta ^{\alpha \beta }-i\ ^{\mathcal{L}}\mathbf{g%
}^{\alpha \beta }.$ The local coordinates on $\mathbf{V}^{2n}$ are
parametrized in the form $u=\{u^{\alpha }\}$ and the local coordinates on $%
T_{u}\mathbf{V}^{2n}$ are labelled $(u,z)=(u^{\alpha },z^{\beta }),$ where $%
z^{\beta }$ are the second order fiber coordinates. We use the formal Wick
product
\begin{equation}
a\circ b\ (z)\doteqdot \exp \left( i\frac{v}{2}\ ^{\mathcal{L}}\mathbf{%
\Lambda }^{\alpha \beta }\frac{\partial ^{2}}{\partial z^{\alpha }\partial
z_{[1]}^{\alpha }}\right) a(z)b(z_{[1]})\mid _{z=z_{[1]}}.  \label{fpr}
\end{equation}%
for two elements $a$ and $b$ defined by formal series of type
\begin{equation}
a(v,z)=\sum\limits_{r\geq 0,|\overbrace{\alpha }|\geq 0}\ a_{r,\overbrace{%
\alpha }}(u)z^{\overbrace{\alpha }}\ v^{r},  \label{formser}
\end{equation}%
where $\overbrace{\alpha }$ is a multi--index, defining the formal Wick
algebra $\mathbf{W}_{u}$ or $u\in \mathbf{V}^{2n}$ associated \ with the
tangent space $T_{u}\mathbf{V}^{2n}$ (a boldface letter emphasizes what we
perform our constructions for spaces provided with N--connection structure).

The fibre product (\ref{fpr}) can be trivially extended to the space of $%
\mathbf{W}$--valued N--adapted differential forms $\ ^{\mathcal{L}}\mathcal{W%
}\otimes \Lambda $ by means of the usual exterior product of the scalar
forms $\Lambda ,$ where $\ ^{\mathcal{L}}\mathcal{W}$ denotes the sheaf of
smooth sections of $\mathbf{W}$ (we put the left label $\mathcal{L}$ in
order to emphasize that the constructions are adapted to the canonical
N--connection structure induced by a regular effective Lagrangian). There is
a standard grading on $\Lambda \mathbf{,}$ denoted $\deg _{a}.$ It is
possible to introduce grading $\deg _{v},\deg _{s},\deg _{a}$ on $\ ^{L}%
\mathcal{W}\otimes \Lambda $ defined on homogeneous elements $v,z^{\alpha },%
\mathbf{e}^{\alpha }$ as follows: $\deg _{v}(v)=1,$ $\deg _{s}(z^{\alpha
})=1,$ $\deg _{a}(\mathbf{e}^{\alpha })=1,$ and all other gradings of the
elements $v,z^{\alpha },\mathbf{e}^{\alpha }$ are set to zero. The product $%
\circ $ from (\ref{fpr}) on $\ ^{L}\mathcal{W}\otimes \mathbf{\Lambda }$ is
bigrated, we write w.r.t the grading $Deg=2\deg _{v}+\deg _{s}$ and the
grading $\deg _{a}.$

The canonical d--connection $\ ^{\mathcal{L}}\widehat{\mathbf{D}}\mathbf{=\{}%
\ ^{\mathcal{L}}\widehat{\mathbf{\Gamma }}\mathbf{_{\alpha \beta }^{\gamma
}\}}$ can be extended to an operator on $\ ^{\mathcal{L}}\mathcal{W}\otimes
\Lambda ,$
\begin{equation}
\ ^{\mathcal{L}}\widehat{\mathbf{D}}\left( a\otimes \lambda \right)
\doteqdot \left( \ ^{\mathcal{L}}\mathbf{e}_{\alpha }(a)-u^{\beta }\ ^{%
\mathcal{L}}\widehat{\mathbf{\Gamma }}\mathbf{_{\alpha \beta }^{\gamma }\ }%
^{z}\mathbf{e}_{\alpha }(a)\right) \otimes (\ ^{\mathcal{L}}\mathbf{e}%
^{\alpha }\wedge \lambda )+a\otimes d\lambda ,  \label{cdcop}
\end{equation}%
where $^{z}\mathbf{e}_{\alpha }$ is a similar to $\ ^{\mathcal{L}}\mathbf{e}%
_{\alpha }$ but depending on $z$--variables (for holonomic second order
fibers, we can take $^{z}\mathbf{e}_{\alpha }=\partial /\partial z^{\alpha
}).$ Using formulas (\ref{fpr}) and (\ref{cdcop}), we can show that $\ \ ^{%
\mathcal{L}}\widehat{\mathbf{D}}$ is a N--adapted $\deg _{a}$--graded
derivation of the distinguished algebra $\left( \ \ ^{\mathcal{L}}\mathcal{W}%
\otimes \mathbf{\Lambda ,\circ }\right) ,$ in brief, one call d--algebra.
This allows us to define on $\ \ ^{\mathcal{L}}\mathcal{W}\otimes \mathbf{%
\Lambda }$ the Fedosov operators $\ \ ^{\mathcal{L}}\delta $ and $\ ^{%
\mathcal{L}}\delta ^{-1}:$
\begin{equation*}
\ ^{\mathcal{L}}\delta (a)=\ ^{\mathcal{L}}\mathbf{e}^{\alpha }\wedge
\mathbf{\ }^{z}\mathbf{e}_{\alpha }(a)\mbox{ and }\ \ ^{\mathcal{L}}\delta
^{-1}(a)=\left\{
\begin{array}{c}
\frac{i}{p+q}z^{\alpha }\ ^{\mathcal{L}}\mathbf{e}_{\alpha }(a),\mbox{ if }%
p+q>0, \\
{\qquad 0},\mbox{ if }p=q=0,%
\end{array}%
\right.
\end{equation*}%
where $a\in \ \ ^{\mathcal{L}}\mathcal{W}\otimes \mathbf{\Lambda }$ is
homogeneous w.r.t. the grading $\deg _{s}$ and $\deg _{a}$ with $\deg
_{s}(a)=p$ and $\deg _{a}(a)=q.$ One holds the formula $a=(\ ^{\mathcal{L}%
}\delta \ ^{\mathcal{L}}\delta ^{-1}+\ \ ^{\mathcal{L}}\delta ^{-1}\ \ ^{%
\mathcal{L}}\delta +\sigma )(a),$ where $a\longmapsto \sigma (a)$ is the
projection on the $(\deg _{s},\deg _{a})$--bihomogeneous part of $a$ of
degree zero, $\deg _{s}(a)=\deg _{a}(a)=0.$ One can be verified that $%
^{L}\delta $ is also a $\deg _{a}$--graded derivation of the d--algebra $%
\left( \ ^{\mathcal{L}}\mathcal{W}\otimes \mathbf{\Lambda ,\circ }\right) .$

The d--connection $\ ^{\mathcal{L}}\widehat{\mathbf{D}}$ on $\mathbf{V}^{2n}$
induces on$\ \ ^{\mathcal{L}}\mathcal{W}\otimes \mathbf{\Lambda }$ the
operators
\begin{eqnarray}
\ ^{\mathcal{L}}\widehat{\mathcal{T}} &\doteqdot &\frac{z^{\gamma }}{2}\ \ ^{%
\mathcal{L}}\theta _{\gamma \tau }\ ^{\mathcal{L}}\widehat{\mathbf{T}}%
_{\alpha \beta }^{\tau }(u)\ ^{\mathcal{L}}\mathbf{e}^{\alpha }\wedge \ ^{%
\mathcal{L}}\mathbf{e}^{\beta },  \label{at1} \\
\ ^{\mathcal{L}}\widehat{\mathcal{R}} &\doteqdot &\frac{z^{\gamma
}z^{\varphi }}{4}\ \ ^{\mathcal{L}}\theta _{\gamma \tau }\ ^{\mathcal{L}}%
\widehat{\mathbf{R}}_{\ \varphi \alpha \beta }^{\tau }(u)\ ^{\mathcal{L}}%
\mathbf{e}^{\alpha }\wedge \ ^{\mathcal{L}}\mathbf{e}^{\beta },  \label{ac1}
\end{eqnarray}%
where the nontrivial coefficients of $\ ^{\mathcal{L}}\widehat{\mathbf{T}}%
_{\alpha \beta }^{\tau }$ and $\ ^{\mathcal{L}}\widehat{\mathbf{R}}_{\
\varphi \alpha \beta }^{\tau }$ are defined completely by formulas (\ref%
{cdtors}) and (\ref{cdcurv}) by introducing the coefficients of $^{\mathcal{L%
}}\widehat{\mathbf{D}}$. For such operators, one hold the formulas $\left[
\mathcal{\ }^{\mathcal{L}}\widehat{\mathbf{D}},\ ^{\mathcal{L}}\delta \right]
=\frac{i}{v}ad_{Wick}(\ ^{\mathcal{L}}\widehat{\mathcal{T}})$ and $\ ^{%
\mathcal{L}}\widehat{\mathbf{D}}^{2}=-\frac{i}{v}ad_{Wick}(\ ^{\mathcal{L}}%
\widehat{\mathcal{R}}),$ where $[\cdot ,\cdot ]$ is the $\deg _{a}$--graded
commutator of endomorphisms of $\ \ ^{\mathcal{L}}\mathcal{W}\otimes \mathbf{%
\Lambda }$ and $ad_{Wick}$ is defined via the $\deg _{a}$--graded commutator
in $\left( \mathcal{\ ^{\mathcal{L}}W}\otimes \mathbf{\Lambda ,\circ }%
\right) .$

\subsection{Fedosov's theorems for effective Lagrange spaces}

We reformulate three theorems (see Theorems 4.1, 4.2 and 4.3 in \cite{vqgfl}%
) generalizing to effective Lagrange spaces the fundamental properties of
Fedosov's d--operators introduced in the pervious section.

\begin{theorem}
\label{th2}Any d--tensor $h_{ij}(x,y)$ and corresponding effective Lagrange
geometry define a flat canonical Fedosov d--connection
\begin{equation*}
\ ^{\mathcal{L}}\widehat{\mathcal{D}}\doteqdot -\ ^{\mathcal{L}}\delta +\ ^{%
\mathcal{L}}\widehat{\mathbf{D}}-\frac{i}{v}ad_{Wick}(r)
\end{equation*}%
satisfying the condition $\ ^{\mathcal{L}}\widehat{\mathcal{D}}^{2}=0,$
where the unique element $r\in $ $\ ^{\mathcal{L}}\mathcal{W}\otimes \mathbf{%
\Lambda ,}$ $\deg _{a}(r)=1,$ $\ ^{\mathcal{L}}\delta ^{-1}r=0,$ solves the
equation
\begin{equation*}
\ ^{\mathcal{L}}\delta r=\ ^{\mathcal{L}}\widehat{\mathcal{T}}+\ ^{\mathcal{L%
}}\widehat{\mathcal{R}}+\ ^{\mathcal{L}}\widehat{\mathbf{D}}r-\frac{i}{v}%
r\circ r
\end{equation*}%
and this element can be computed recursively with respect to the total
degree $Deg$ as follows:%
\begin{eqnarray*}
r^{(0)} &=&r^{(1)}=0,\ r^{(2)} =\ ^{\mathcal{L}}\delta ^{-1}\ ^{\mathcal{L}}%
\widehat{\mathcal{T}}, \mbox{ \ and,\ for\ } k\geq 1, \\
r^{(3)} &=&\ \ ^{\mathcal{L}}\delta ^{-1}\left( \ ^{\mathcal{L}}\widehat{%
\mathcal{R}}+\ \ ^{\mathcal{L}}\widehat{\mathbf{D}}r^{(2)}-\frac{i}{v}%
r^{(2)}\circ r^{(2)}\right) , \\
r^{(k+3)} &=&\ \ ^{\mathcal{L}}\delta ^{-1}\left( \ ^{\mathcal{L}}\widehat{%
\mathbf{D}}r^{(k+2)}-\frac{i}{v}\sum\limits_{l=0}^{k}r^{(l+2)}\circ
r^{(l+2)}\right),
\end{eqnarray*}%
where we denoted the $Deg$--homogeneous component of degree $k$ of an
element $a\in $ $\ \ ^{\mathcal{L}}\mathcal{W}\otimes \mathbf{\Lambda }$ by $%
a^{(k)}.$
\end{theorem}

The next theorem gives a rule how to define and compute the star product
induced by a generalized d--tensor $h_{ij}$:

\begin{theorem}
\label{th3}A star--product on the canonical effective almost K\"{a}%
hler--La\-gran\-ge space is defined on $C^{\infty }(\mathbf{V}^{2n})[[v]]$
by formula
\begin{equation*}
\ ^{1}f\ast \ ^{2}f\doteqdot \sigma (\tau (\ ^{1}f))\circ \sigma (\tau (\
^{2}f)),
\end{equation*}%
where the projection $\sigma :\ ^{\mathcal{L}}\mathcal{W}_{\widehat{\mathcal{%
D}}}\rightarrow C^{\infty }(\mathbf{V}^{2n})[[v]]$ onto the part of $\deg
_{s}$--degree zero is a bijection and the inverse map $\tau :C^{\infty }(%
\mathbf{V}^{2n})[[v]]\rightarrow \ ^{\mathcal{L}}\mathcal{W}_{\widehat{%
\mathcal{D}}}$ can be calculated recursively w.r..t the total degree $Deg,$%
\begin{eqnarray*}
\tau (f)^{(0)} &=&f, \mbox{ \ and,\ for\ } k\geq 0,  \\
\tau (f)^{(k+1)} &=& \ ^{\mathcal{L}}\delta ^{-1}\left( \ ^{\mathcal{L}}%
\widehat{\mathbf{D}}\tau (f)^{(k)}-\frac{i}{v}\sum%
\limits_{l=0}^{k}ad_{Wick}(r^{(l+2)})(\tau (f)^{(k-l)})\right).
\end{eqnarray*}
\end{theorem}

We denote by $\ ^{f}\xi $ the corresponding Hamiltonian vector field
corresponding to a function $f\in C^{\infty }(TM)$ on space $(\mathbf{V}%
^{2n},\ ^{\mathcal{L}}\theta )$ and consider the antisymmetric part $\
^{-}C(\ ^{1}f,\ ^{2}f)\ \doteqdot \frac{1}{2}\left( C(\ ^{1}f,\ ^{2}f)-C(\
^{2}f,\ ^{1}f)\right)$ of bilinear operator $C(\ ^{1}f,\ ^{2}f).$ A
star--product (\ref{starp}) is normalized if $\ _{1}C(\ ^{1}f,\ ^{2}f)=\frac{%
i}{2}\{\ ^{1}f,\ ^{2}f\},$  where $\{\cdot ,\cdot \}$ is the Poisson
bracket. For normalized $\ast $ the bilinear operator $\ _{2}^{-}C$ is a de
Rham--Chevalley 2--cocycle. There is a unique closed 2--form $\ ^{\mathcal{L}%
}\varkappa ,$ in our case induced by a regular effective Lagrangian $%
\mathcal{L},$ such that%
\begin{equation}
\ _{2}C(\ ^{1}f,\ ^{2}f)=\frac{1}{2}\ ^{\mathcal{L}}\varkappa (\ ^{f_{1}}\xi
,\ ^{f_{2}}\xi )  \label{c2}
\end{equation}%
for all $\ ^{1}f,\ ^{2}f\in C^{\infty }(\mathbf{V}^{2n}).$ The class $c_{0}$
of a normalized star--product $\ast $ is defined as the equivalence class $%
c_{0}(\ast )\doteqdot \lbrack \ ^{\mathcal{L}}\varkappa ].$

A straightforward computation of $\ _{2}C$ from (\ref{c2}), using statements
of Theorem \ref{th2}, results in a proof of

\begin{lemma}
\label{lem1}The unique 2--form can be computed
\begin{eqnarray*}
\ ^{\mathcal{L}}\varkappa &=&-\frac{i}{8}\mathbf{J}_{\tau }^{\ \alpha
^{\prime }}\ ^{\mathcal{L}}\widehat{\mathbf{R}}_{\ \alpha ^{\prime }\alpha
\beta }^{\tau }\ ^{\mathcal{L}}\mathbf{e}^{\alpha }\wedge \ ^{\mathcal{L}}%
\mathbf{e}^{\beta }-i\ \ ^{\mathcal{L}}\lambda , \\
\ ^{\mathcal{L}}\lambda &=&d\ \ ^{\mathcal{L}}\mu ,\ \ \ ^{\mathcal{L}}\mu =%
\frac{1}{6}\mathbf{J}_{\tau }^{\ \alpha ^{\prime }}\ ^{\mathcal{L}}\widehat{%
\mathbf{T}}_{\ \alpha ^{\prime }\beta }^{\tau }\ ^{\mathcal{L}}\mathbf{e}%
^{\beta }.
\end{eqnarray*}
\end{lemma}

Next, we define the canonical class $\varepsilon $ for $\ ^{N}T\mathbf{V}%
^{2n}=h\mathbf{V}^{2n}\oplus v\mathbf{V}^{2n}$ (\ref{whitney}) with the left
label stating a N--connection structure $\mathbf{N}.$ The distinguished
complexification of such second order tangent bundles is introduced in the
form $T_{\mathbb{C}}\left( \ ^{N}T\mathbf{V}^{2n}\right) =T_{\mathbb{C}%
}\left( h\mathbf{V}^{2n}\right) \oplus T_{\mathbb{C}}\left( v\mathbf{V}%
^{2n}\right).$  In this case, the class $\ ^{N}\varepsilon $ is the first
Chern class of the distributions $T_{\mathbb{C}}^{\prime }\left( \ ^{N}T%
\mathbf{V}^{2n}\right) =T_{\mathbb{C}}^{\prime }\left( h\mathbf{V}%
^{2n}\right) \oplus T_{\mathbb{C}}^{\prime }\left( v\mathbf{V}^{2n}\right)$
of couples of vectors of type $(1,0)$ both for the h-- and v--parts.

Our aim is to calculate the canonical class $^{\mathcal{L}}\varepsilon $
(the label $\mathcal{L}$ is for the constructions canonically defined by a
regular effective Lagrangian $\mathcal{L}).$ We take the canonical
d--connection $\ ^{\mathcal{L}}\widehat{\mathbf{D}}$ that it was used for
constructing $\ast $ and considers h- and v--projections $h\Pi =\frac{1}{2}%
(Id_{h}-iJ_{h})$ and $v\Pi  =\frac{1}{2}(Id_{v}-iJ_{v}),$ where $Id_{h}$ and
$Id_{v}$ are respective identity operators and $J_{h}$ and $J_{v}$ are
almost complex operators, which are projection operators onto corresponding $%
(1,0)$--subspaces. The matrix $\left( h\Pi ,v\Pi \right) \ ^{\mathcal{L}}%
\widehat{\mathbf{R}}\left( h\Pi ,v\Pi \right) ^{T},$ where $(...)^{T}$ means
transposition, is the curvature matrix of the N--adapted restriction of the
connection $\ \ ^{\mathcal{L}}\widehat{\mathbf{D}}$ to $T_{\mathbb{C}%
}^{\prime }\left( \ ^{N}T\mathbf{V}^{2n}\right) .$ Now, we can compute the
Chern--Weyl form
\begin{eqnarray*}
\ ^{\mathcal{L}}\gamma &=&-iTr\left[ \left( h\Pi ,v\Pi \right) \widehat{%
\mathbf{R}}\left( h\Pi ,v\Pi \right) ^{T}\right] =-iTr\left[ \left( h\Pi
,v\Pi \right) \ ^{\mathcal{L}}\widehat{\mathbf{R}}\right] \\
&=&-\frac{1}{4}\mathbf{J}_{\tau }^{\ \alpha ^{\prime }}\ ^{\mathcal{L}}%
\widehat{\mathbf{R}}_{\ \alpha ^{\prime }\alpha \beta }^{\tau }\ ^{\mathcal{L%
}}\mathbf{e}^{\alpha }\wedge \ ^{\mathcal{L}}\mathbf{e}^{\beta }
\end{eqnarray*}%
to be closed. By definition, the canonical class is $\ ^{\mathcal{L}%
}\varepsilon \doteqdot \lbrack \ \ ^{\mathcal{L}}\gamma ].$

\begin{theorem}
The zero--degree cohomology coefficient $c_{0}(\ast )$ for the almost K\"{a}%
hler model of effective Lagrange space defined by a d--tensor $h_{ij}(x,y)$
is computed
\begin{equation*}
c_{0}(\ast )=-(1/2i)\ \ ^{\mathcal{L}}\varepsilon .
\end{equation*}
\end{theorem}

For a prescribed regular effective Lagrangian, such values can be defined,
for instance, by an exact solution in Einstein gravity on a manifold $M$, $%
\dim M=n,$ lifted canonically to $TM,$ or extended to an effective Lagrange
model on $\mathbf{V}^{2n}.$

Finally, we note that theorems from this section can be re--defined for
nonholonomic  deformations on  Einstein manifolds, see \cite%
{vqgd}.

\end{document}